\documentclass[prl,twocolumn,showpacs,preprintnumbers,reprint,amsmath, amssymb,superscriptaddress]{revtex4-1} % {{{

\usepackage{wasysym,helvet}
\usepackage{graphicx} % Include figure files
\usepackage{dcolumn}  % Align table columns on decimal point
\usepackage{bm}       % bold math
\usepackage{times,mathptmx}
\usepackage{color}

\usepackage{soul}

\graphicspath{{figuras/}}

\begin{document}
\newif\ifnocolor
\nocolortrue

\newcommand{\bea}{\begin{eqnarray}}
\newcommand{\eea}{ \end{eqnarray}}
\newcommand{\bit}{\begin{itemize}}
\newcommand{\eit}{ \end{itemize}}

\newcommand{\be}{\begin{equation}}
\newcommand{\ee}{\end{equation}}
\newcommand{\ra}{\rangle}
\newcommand{\la}{\langle}
\newcommand{\U}{\widetilde{U}}
\newcommand{\brac}[1]{\langle #1|}
\newcommand{\bra}[1]{\langle #1}
\newcommand{\ket}[1]{|#1\rangle}
\newcommand{\ktimes}{\rangle\! \langle}
\newcommand{\op}[2]{|#1\ktimes #2|}
\newcommand{\opoo}{\op{0}{0}}
\newcommand{\opoi}{\op{0}{1}}
\newcommand{\opio}{\op{1}{0}}
\newcommand{\opii}{\op{1}{1}}
\newcommand{\rhoq}{\rho_{\rm q}}
\newcommand{\rhoe}{\rho_{\rm env}}
\newcommand{\env}{{\rm env}}
\newcommand{\qubit}{{\rm qubit}}
\newcommand{\Mp}{{\cal M}_{\rm p}}
\newcommand{\Mm}{{\cal M}_{\rm m}}
\newcommand{\NPC}{\xi}
\newcommand{\ANPC}{\langle \xi\rangle}
% \newrgbcolor{fbkblue}{0.2304 0.3476  0.5937}
% \definecolor{fbkblue}{rgb}{0.2304 0.3476  0.5937}
% \definecolor{pksblue}{rgb}{0.137 0.298 0.513}
% \newrgbcolor{fbkblue}{0.2304 0.3476  0.5937}
% \newrgbcolor{pksblue}{0.137 0.298 0.513}
% \newrgbcolor{verde}{0.267 0.637 0.492}  %%#76 163 126
% \newrgbcolor{verde2}{0.168 0.582 0.543} %%43 149 139
% \newcommand{\fbk}[1]{{\color{fbkblue} #1}}

%----------new commands-------------

\def\bracket#1#2{{\langle#1|#2\rangle}}
\def\inner#1#2{{\langle#1|#2\rangle}}
\def\expect#1{{\langle#1\rangle}}
\def\e{{\rm e}}
\def\proj{{\hat{\cal P} }}
\def\tr{{\rm Tr}}
\def\H{{\hat H}}
\def\Hdag{{\hat H}^\dagger}
\def\Lop{{\cal L}}
\def\Ehat{{\hat E}}
\def\Edag{{\hat E}^\dagger}
\def\Shat{\hat{S}}
\def\Sdag{{\hat S}^\dagger}
\def\Ahat{{\hat A}}
\def\Adag{{\hat A}^\dagger}
\def\U{{\hat U}}
\def\Udag{{\hat U}^\dagger}
\def\Zhat{{\hat Z}}
\def\Phat{{\hat P}}
\def\Op{{\hat O}}
\def\id{{\hat I}}
\def\x{{\hat x}}
\def\P{{\hat P}}	
\def\Px{\proj_x}
\def\Pr{\proj_{R}}
\def\Pl{\proj_{L}}
\def\ODR{f_{_{\rm DR}}(t)}
\def\ODRn{O_{_{\rm DR}}(n)}
\newcommand{\equa}[1]{Eq.~(\ref{#1})}
\newcommand{\comm}[1]{{\color{googB}[#1]}}
\newcommand{\comp}[1]{{\bf #1}}
\newcommand{\mcH}{\mathcal{H}}
\renewcommand{\e}{\text{env}}
\newcommand{\s}{{\rm sys}}
\newcommand{\eref}[1]{Eq.~(\ref{#1})}
\newcommand{\Eref}[1]{Eq.~(\ref{#1})}
\newcommand{\Sdec}{S_{\rm dec}} 		%% entro deco
\newcommand{\SD}{S_{\rm D}} 			%% DIAGONAL entro
\newcommand{\lambdac}{\lambda_{\rm c}}	%% critical lambda
\newcommand{\IQS}{\textsf{IQS} }
\newcommand{\nmax}{n_{\rm max}}

\def\co#1{{\color{red}\textst{#1}}}
\newcommand{\aug}[1]{#1}
\definecolor{verde}{rgb}{0.458,0.765,0.25}
 \definecolor{googB}{rgb}{0.285,0.539,0.949}
 \definecolor{fbkblue}{rgb}{0.2304, 0.3476,  0.5937}
\newcommand{\nach}[1]{{\color{verde} #1}}

%----------end new commands-------------
% \newrgbcolor{googB}{0.285 0.539 0.949}
% \newrgbcolor{googR}{0.832 0.285 0.215}
%---------end new colors ---------------------
%%%%%%%%%%%%
\title{Relaxation of isolated quantum systems beyond chaos} 
%%%
\author{Ignacio Garc\'{\i}a-Mata}
\affiliation{Instituto de Investigaciones F\'isicas de Mar del Plata (IFIMAR, CONICET), 
Universidad Nacional de Mar del Plata, Mar del Plata, Argentina.}
\email{i.garcia-mata@conicet.gov.ar}
\affiliation{Consejo Nacional de Investigaciones Cient\'ificas y Tecnol\'ogicas (CONICET), Argentina}
\author{Augusto J. Roncaglia}
\affiliation{\mbox{Departamento de F\'{\i}sica ``J. J. Giambiagi" and IFIBA, 
             FCEyN, Universidad de Buenos Aires, 1428 Buenos Aires, Argentina}}                         
\author{Diego A. Wisniacki}
\affiliation{\mbox{Departamento de F\'{\i}sica ``J. J. Giambiagi" and IFIBA, 
             FCEyN, Universidad de Buenos Aires, 1428 Buenos Aires, Argentina}}
%%%%%%%%%%%%
\date{12th August 2014 -- submitted}
%%%%%%%%%%%%%%

\begin{abstract} 
In classical statistical mechanics there is a clear correlation between relaxation to equilibrium and chaos. 
In contrast, for isolated quantum systems this relation is  $-$ to say the least $-$ fuzzy.   In this work we try to unveil
the intricate relation between the relaxation process and the transition from integrability to chaos. We study the approach to equilibrium in  two different many body quantum systems 
that can be parametrically tuned from regular to chaotic. 
We show that a universal relation between relaxation and 
delocalization of the initial state in the perturbed basis can be established regardless of the chaotic nature of system.
\end{abstract}
%%%%%%%%%%%%% 03.65.Yz, 03.67.Mn, 05.45.Mt
%--
%05.45.Pq	Numerical simulations of chaotic systems
%05.45.Mt	Quantum chaos; semiclassical methods
%03.65.Sq	Semiclassical theories and applications
%03.65.Yz	Decoherence; open systems; quantum statistical methods 
%	      		(see also 03.67.Pp in quantum information; for decoherence in 
%	     		 Bose-Einstein condensates, see 03.75.Gg)
% 05.70.Ln 	Nonequilibrium and irreversible thermodynamics
% 05.30.-d 	Quantum statistical mechanics
% 42.50.Ct  Quantum description of interaction of light and matter; related experiments
% 64.70.Tg	Quantum phase transitions (for quantum Hall effects aspects, see 73.43.Nq in electronic structure of surfaces, interfaces, thin films, and low dimensional structures)
%03.67.Lx Quantum computation architectures and implementations
 \maketitle

The second law is the cornerstone upon which the strength of 
thermodynamics
lies \cite{Callen}. It states that during a process, the entropy of an isolated system should increase.
A process delivers an initial equilibrium state to another one, thus it is assumed that during the evolution
for sufficiently long time an equilibrium state is achieved. In statistical  mechanics  
 relaxation can be formalized by the concept of weak mixing \cite{arnold}, a property that
 is accomplished by chaotic systems. 
 
In quantum mechanics the situation is more subtle. First, there is no
straightforward translation of the concept of classical chaos to the quantum realm.
The definition of classical chaos depends on 
exponential separation of phase space trajectories and mixing \cite{ChaoticScattering}. 
Although these two notions in quantum systems are devoid of meaning,
there are certainly other ways to define quantum chaos, mainly through 
spectral statistics \cite{bohigas} and properties of eigenfunctions \cite{haake,*stockmann}. 
The second reason is that the straightforward generalization of
classical entropy to quantum physics, the von Neumann entropy $S_{\rm vN}=-{\rm Tr}(\rho\ln\rho)$,  
is preserved for any process in closed systems. %% 
Thus it does not comply with the second law for systems out of equilibrium. 
For this reason alternative definitions of entropy have been proposed. One good candidate is the
diagonal entropy (d-entropy) \cite{Polkov2011,Santos2011}, defined as 
%%%%%%%%%%%%%%%%%%%%%%
\begin{equation}
\SD=-\sum_n\rho_{nn}\ln\rho_{nn},
\end{equation}
%%%%%%%%%%%%%%%%%%%%%%
where $\rho_{nn}$
are the diagonal elements of $\rho$ in the energy eigenbasis. 
It is the Shannon entropy of the probability distribution corresponding to the energy
measurement. If the density matrix is a convex combination of energy eigenstates, i.e., for stationary states, 
the d-entropy 
coincides with the $S_{\rm vN}$. On the other hand, $\SD$ 
increases for systems out of equilibrium, 
and satisfies most of the requirements of a thermodynamic entropy \cite{Polkov2011,Santos2011}.

The goal of this communication is to elucidate the  approach to equilibrium of isolated quantum systems whose dynamics is governed by a Hamiltonian that can be tuned from integrable to chaotic. Equilibration \cite{Tasaki1998,*Reimann2008,*Reimann2012,Linden2009} is a less restrictive property but which is (generally) deemed necessary for thermalization, i.e., the study of how isolated quantum systems can relax to states that can be described by usual statistical mechanics
\cite{Jensen85,*Deutsch91,*Srednicki1994,*Rigol2008,*Rigol2009,*Biroli2010,Santos2011,*Ponomarev2011,*GogolinMullerEisert2011,*Ikeda2011,*Riera2012,*Rigol2012}. Here, 
we consider relaxation in quenched dynamics, where a system is perturbed by
a sudden change in the Hamiltonian. The process of relaxation is studied by considering the evolution of the  Shannon entropy 
in the initial equilibrium basis. This is equivalent to considering 
the evolution of the diagonal entropy for
a cyclic process whereby the original Hamiltonian is 
quenched at some initial time, then the system is left to evolve unitarily and finally 
a reversion of the original quench is applied. 
We consider that the  quench is implemented by a sudden change of the (chaos) tuning parameter.
Relaxation is then characterized by increasing entropy and vanishing fluctuations.
Within such framework, extensive numerical simulations were done using 
two different many-body systems: the paradigmatic Dicke model \cite{Dicke54} and  
a spin chain with nearest-neighbor and next-nearest-neighbor couplings. 
At equilibrium, the d-entropy becomes (approximately) constant and the fluctuations of the d-entropy tend to be negligible.
Although
a one to one correlation between chaos and relaxation is to be expected, 
for initial states corresponding to large energy eigenstates, we observe that 
relaxation is achieved for values below the transition
transition from integrable to chaotic. On the other hand, at equilibrium  
the initial state spreads over the perturbed basis, thus becoming increasingly delocalized \cite{Santos2012}. %% 
Therefore,
delocalization (or quantum ergodicity \cite{PolkovnikovRMP2011}),  besides non-integrability,  is a key 
feature for a system to reach quantum equilibrium.  
We show -- numerically -- that there is a universal relation linking the d-entropy at equilibrium with the
inverse participation ratio, which is a measure of the localization properties. 

%%%%%%%%%%%%%%%%%%%%%%%%%%%%%%%%%%%%%%%%%%%%%%%%%%%%%%%%%%%%%%%%%
%%\section{SETUP}
%%%%%%%%%%%%%%%%%%%%%%%%%%%
We consider the following  process.
Initially the system is at state $\rho_0$ and the dynamics is given by a Hamiltonian $H$.
At time
$t=0$ a quench is applied and the system evolves unitarily with the new (time-independent) 
Hamiltonian $H'$. Finally
at time $t=\tau$ another quench changes the Hamiltonian to $H''$. For simplicity, we consider a cyclic operation: $H''=H$.  
The state of the system at time $\tau$ is $\rho(\tau)=e^{-iH'\tau}\rho_0 e^{iH'\tau}$, where we picked $\rho_0=\ket{n_0}\brac{n_0}$, with $\ket{n_0}$ an eigenstate of $H$.  
For time-independent Hamiltonians the d-entropy is constant in time, thus in this case it will only depend on the time of the final quench $\tau$.  
Given that the Hamiltonian for $t\ge\tau$ is $H$, the d-entropy at time $\tau$ is 
\begin{equation}
\SD(\tau)=-\sum_n C_n(\tau)\ln C_n(\tau),
\end{equation}
where $C_n(\tau)=\brac{n}\rho(\tau)\ket{n}=|\bra{n}|e^{-iH' \tau}\ket{n_0}|^2$, and $\ket{n}$ is and an 
eigenstate of $H$.
%%%%
The d-entropy satisfies the second law for typical operational times, that is,
after the quench it grows  and after an equilibration time scale it stabilizes to a constant value.
Since the d-entropy is a non-linear function of the density matrix, its time average is not equal to the 
d-entropy of the time averaged state $\Sdec$. If the time-averaged state is $\rho_{\rm dec}=\overline{\rho(\tau)}$,
where  $\overline{f(\tau) }\equiv \lim_{T\rightarrow \infty} T^{-1}\int_0^T d\tau f (\tau)$, then
$\Sdec = -\sum_n\mu_n \ln \mu_n$ with $\mu_n=\brac{n}\rho_{\rm dec}\ket n$.
Recently it was conjectured  \cite{Ueda2013} that the relaxation to equilibrium is reflected in the following
sub-extensive correction to the time-average of the d-entropy
\begin{equation}
\label{subext}
\Sdec - \overline{\SD(\tau)}\leq 1-\gamma,
\end{equation}
where $\gamma = 0.5772\ldots$ is Euler's constant, provided the initial state is pure. 
As a consequence, the equilibrium condition will also be reflected in 
the fluctuations of  $\Sdec - \overline{\SD(\tau)}$, which should decrease to a minimum.
To shed light on the relation between the transition from integrability to chaos and 
the relaxation process, we study the d-entropy after a quench process is implemented on quantum systems 
where the interaction strength plays the role of integrability parameter.

We start with the paradigmatic Dicke model (DM)
\cite{Dicke54}.
It is especially known for its quantum phase transition to a superradiant phase \cite{HeppLieb1973} that
has been observed recently with a superfluid gas in an optical cavity \cite{dicke-exp}.
The single mode DM  describes the (dipole) interaction between an ensemble of $N$ two-level atoms 
with level splitting $\omega_0$ and a single mode of a bosonic field of frequency $\omega$:
\begin{equation}
\label{Hdicke}
H(\lambda)=\omega_0 J_z+\omega a^\dagger a+\frac{\lambda}{\sqrt{2 j}}
(a^\dagger+a)(J_+ + J_-),
\end{equation}
where $\lambda$ is the coupling constant. Here $J_z$ and $J_\pm$ are collective angular momentum operators for a 
pseudospin of length $j=N/2$, and $a$ ($a^\dagger$) are the bosonic annihilation (creation) operators of the field. In the 
thermodynamic limit ($N\to \infty$) 
there is a superradiant phase transition \cite{HeppLieb1973} at $\lambdac=\sqrt{\omega_0\omega}/2$. For finite $N$ 
there is also a transition at $\lambda\approx\lambdac$ from quasi-integrability, where level spacing statistics is 
Poissonian, to quantum chaos, with typical Wigner-Dyson distribution \cite{EmaryBrandes2003,*EmaryBrandesPRE2003}. 
Interestingly, the chaotic behavior could also be verified using a semiclassical model \cite{Altland2012,*Altland2014}.
We consider $\omega=\omega_0=1$ so that $\lambdac=0.5$ and $\hbar=1$. 
The Dicke Hamiltonian is invariant under parity transformations so we will constrain our calculations 
to the even subspace.
%%%%%%%%%%%%%%%%%%%%%%%%%%%%%%%%%%%%%%%%%%%%%%%
%{[Resultados]}
%%%%%%%%%%%%%%%
%%			FIGURA
%%%%%%%%%%%%%%%%%%%%%%%
\begin{figure}[htb]
\includegraphics[width=\linewidth]{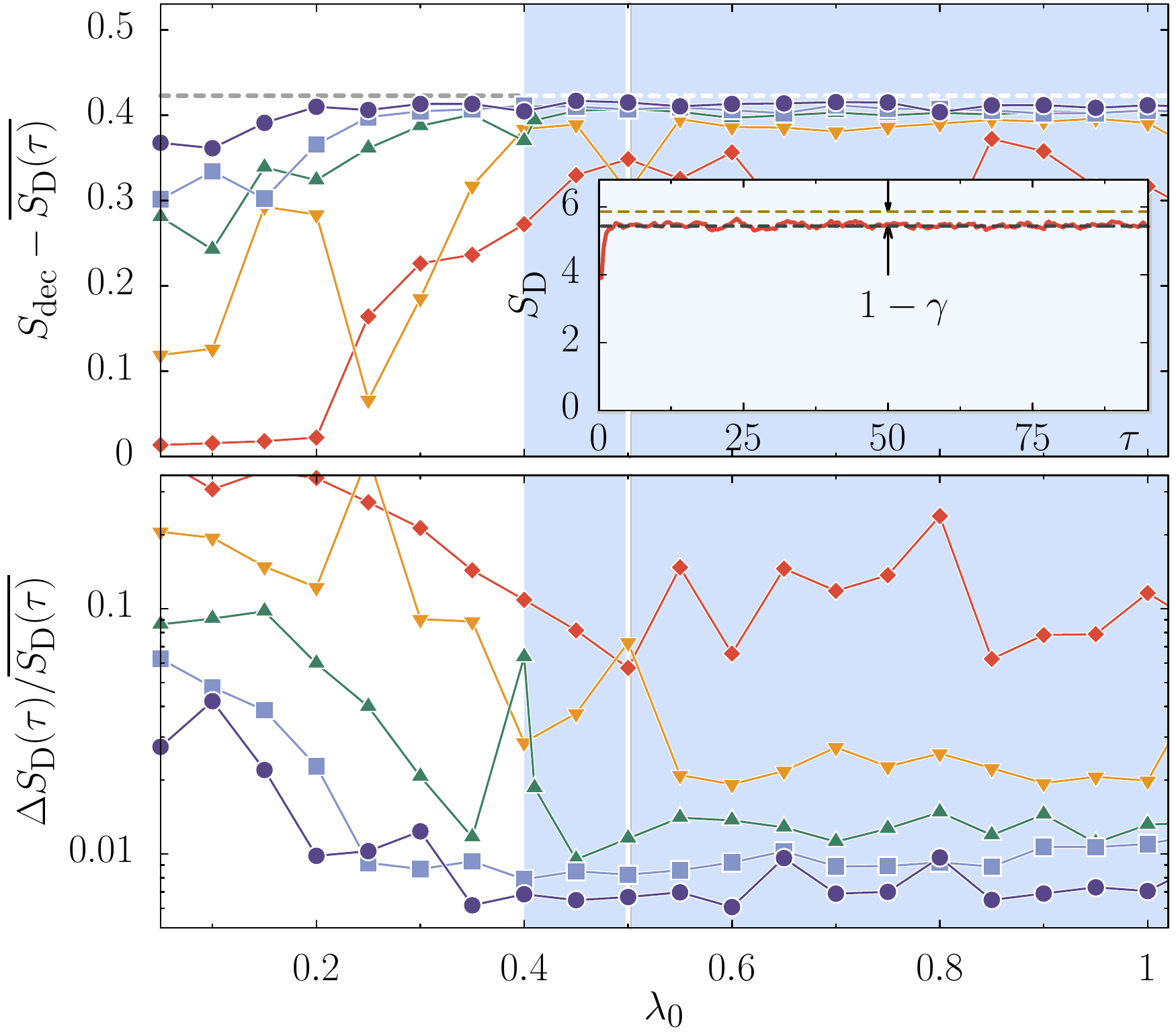} %{fock_Dlam01} %
\caption{\label{fig:lambda} 
(Color online) Top: 
$\Sdec-\overline{\SD(\tau)}$. Bottom: Averaged fluctuations of $\SD(\tau)$ as a function of $\lambda_0$, with 
$\delta \lambda=0.1$.  The vertical (solid) line marks the transition to chaotic behavior at $\lambdac=0.5$. The shaded area 
marks the region where at least one Hamiltonian [$H(\lambda_0)$ or $H(\lambda_0+\delta\lambda)$] corresponds to chaotic 
dynamics. The horizontal dashed (top panel) line marks 
universal value $1-\gamma\approx0.423$. The symbols correspond to different initial states $\ket{n}$ ($H\ket{n}=E_n\ket{n}$):
({\large $\diamond$}) $\ket{10}$; ({\large$\triangledown$}) $\ket{100}$; 
($\triangle$) $\ket{500}$; ($\square$) $\ket{1000}$; ({\Large $\circ$}) $\ket{2000}$. 
Inset: The dashed green line on top marks $\Sdec$ while the dashed gray line marks $\Sdec-1+\gamma$. DM with $J=20$, $
\lambda=0.65$ ($\nmax=250$) energy $E_{501}$. The arrows indicate the distance 
$\Sdec-\SD(\tau)\approx 1-\gamma=0.4228\ldots$ .} 
\end{figure}
%%%%%%%%%%%%%%%%%%%%%%%%

We consider the behavior of the d-entropy 
%\co{in the relaxation by a cyclic process} 
for different quenches, 
where an initial Hamiltonian $H=H(\lambda_0)$ is perturbed by $H'=H(\lambda_0+\delta\lambda)$, 
where $\delta \lambda$ is the quench amplitude.
We did straightforward diagonalization 
%\co{of $H(\lambda_0)$} 
in the Fock basis (taking parity into account). 
The phonon basis was truncated
%\co{We truncated the phonon basis} 
at a value $\nmax\sim 250$.
The typical behavior of the d-entropy as a function of $\tau$ (for $\lambda_0>\lambda_c$) can be seen in the inset of 
Fig.~\ref{fig:lambda}, for the DM ($\lambda_0=0.65,\ \delta\lambda=0.1$). After a short period of time the d-entropy
settles approximately to a constant value. 
The dashed line corresponds to $\Sdec$ and the difference marked by the arrows is $1-\gamma$, in correct accordance with Eq.~(\ref{subext}).  
%is exactly given by Eq.~(\ref{subext}).

We now systematically change the parameter $\lambda_0$, leaving fixed $\delta \lambda=0.1$ 
\footnote{We have checked that for other values of $\delta \lambda$ the results are equivalent.} and compute $\SD(\tau)$
for different initial states $\ket{n}$ (where $H\ket{n}=E_n\ket{n}$). 
Then we evaluated  the time-average $\overline{\SD(\tau)}=\sum_{\tau}^{\tau+\delta\tau}\SD(\tau)/n_{\rm steps}$ 
(where $n_{\rm steps}$ is the number of equally spaced time steps) and variance $\Delta \SD(\tau)$ 
of the time average. 
The  time window is defined by $[\tau,\tau+\delta\tau]$, where $\tau$ is much larger than the short time scale,   
$\delta \tau\sim 250$, and is subdivided into 1000 time steps.
%AUNQUE QUIZAS SACARIA ESTOS DETALLES.
%%%%%
In Fig.~\ref{fig:lambda} (top) $\Sdec-\overline{\SD(\tau)}$ is displayed as a function of $\lambda_0$.
Our data show that as the strength of the interaction term increases, rendering the system more chaotic, 
$\Sdec-\overline{\SD(\tau)}$ tends to the bound given by Eq. \eqref{subext}.
In addition, as the energy of the initial state increases, $\Sdec-\overline{\SD(\tau)}$ attains a value closer to the bound even
for $\lambda_0<0.5$ where the system is not chaotic.
In Fig.~\ref{fig:lambda} (bottom) we show the fluctuations $\Delta \SD/\overline{\SD}$,
where $(\Delta \SD)^2=\overline{\SD(\tau)^2}-(\overline{\SD(\tau)})^2$, for the same initial states.
It is remarkable the relation between  $\Sdec-\overline{\SD(\tau)}$ and its fluctuations:
as $\Sdec-\overline{\SD(\tau)}$ gets closer to the bound the fluctuations are smaller.
Thus, for initial eigenstates with low energy equilibration is hardly achieved.
 
%%%%%%%%%%%%%%%%%%%%%%%
\begin{figure}%[htb]
\includegraphics[width=\linewidth]{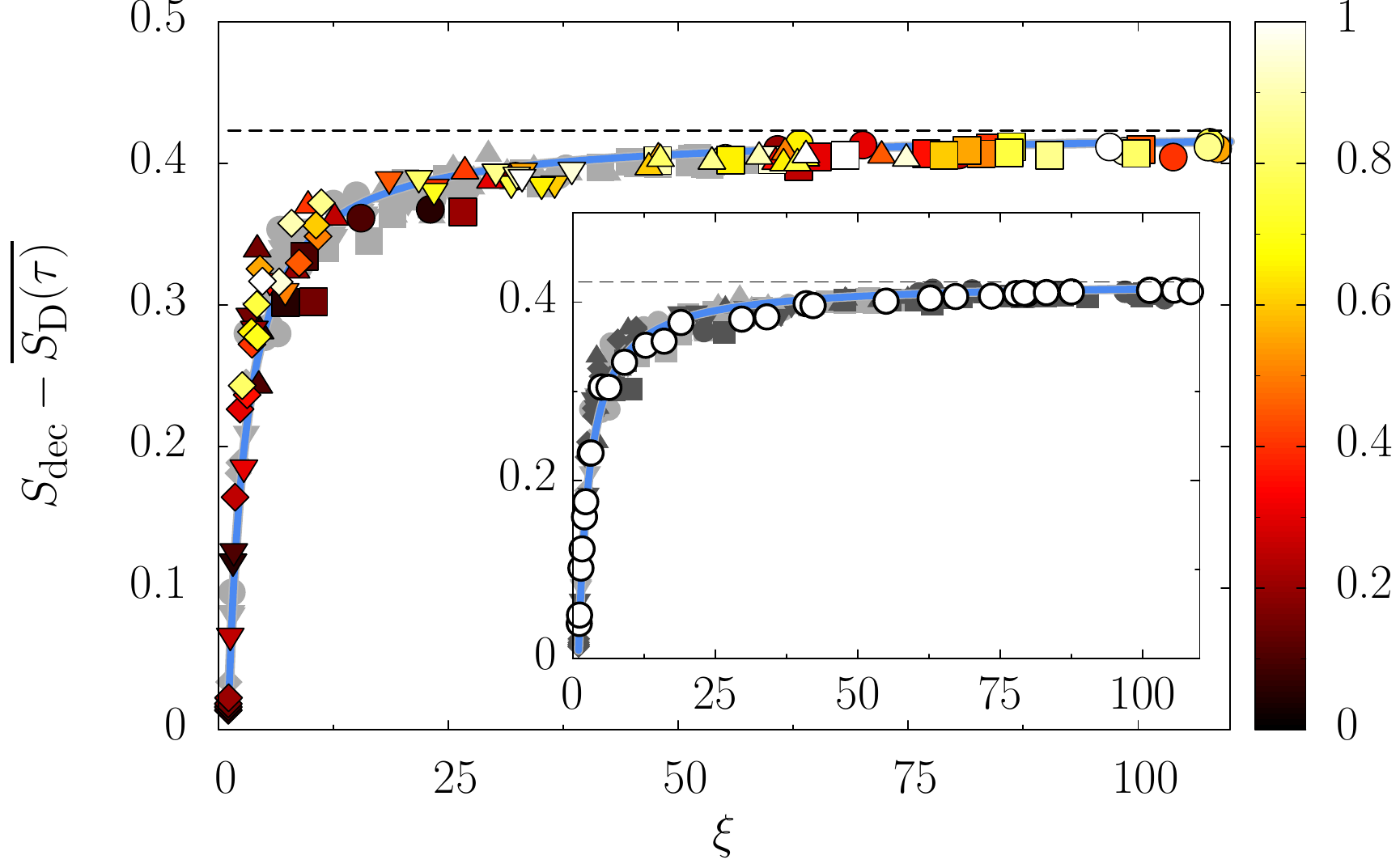} %
\caption{\label{fig:SdeIPR} 
(Color online) $\Sdec-\overline{\SD(\tau)}$ as a function of $\NPC$. Main panel: 
DM with $j=20$, $\lambda\le 1.0$ and $\delta \lambda=0.1$, and $\nmax=250$. The color-box 
indicates the values of $\lambda$.  The dashed, horizontal line marks the 
value $1-\gamma=0.4228\ldots\, $. The 
symbols correspond to different initial states: ({\large $\diamond$}) $\ket{10}$; ({\large$
\triangledown$}) $\ket{100}$; ($\triangle$) $\ket{500}$; ($\square$) $\ket{1000}$; ({\Large $
\circ$}) $\ket{2000}$; light-gray symbols correspond to results for $j=10$. 
The solid/blue curve is the approximating function $(1-
\gamma)(\NPC-1)/(\NPC+1)$.
Inset:  
the white ({\Large$\circ$}) correspond to results for the spin model with $\mu=0.5$.
%Same as in the main panel but for the spin model ({\Large$\circ$}) with $\mu=0.5$, $\delta\lambda ...$. 
The gray symbols correspond to the DM results shown in the main panel (light gray $j=10$ ; dark gray $j=20$).
} 
\end{figure}
%%%%%%%%%%%%%%%%%%%%%%%%

Results in Fig.~\ref{fig:lambda} suggest that there is a deep connection between three quantities:
$\Sdec- \overline\SD$, the energy of the initial state, and $\lambda_0$.
Equilibration of $\SD$ as $\lambda_0$ increases is expected since chaoticity also increases with $\lambda_0$. 
However, the behavior observed for initial eigenstates far from the low energy region
in the quasi-integrable regime is more unusual 
compared with classical systems. Interestingly, it is known that for quantum systems the complexity of the
eigenstates also provides a mechanics for relaxation \cite{Linden2009}.
Indeed,  we will show that there is a quantity 
that connects the equilibrium properties of $\SD$ with the initial state and dynamics,
namely, the inverse participation ratio (IPR). The IPR of an initial state $\ket{n(\lambda)}$ 
in the perturbed basis is 
%\begin{equation}
$\xi=\left(\sum_m | \inner{n(\lambda)}{m(\lambda_0+\delta \lambda)}|^4\right)^{-1}$.
%\end{equation}
This quantity estimates the number of perturbed states contributing to a given unperturbed state and has been widely used, e.g., to describe (de-)localization in relation to chaos \cite{Fyodorov1995,Jacquod1995,Georgeot1997}, Anderson transition in Fock space\cite{Berkovits1998}, analyzing the structure of a real quantum chaotic system -- the cessium atom--\cite{Flambaum1994}, and entropy production for chaotic systems \cite{Flambaum2001}.  
%%%%%%%%%%%%%%%%%%%%%%%% Sdec_Smed_todos.pdf
\begin{figure}%[htb]
\includegraphics[width=\linewidth]{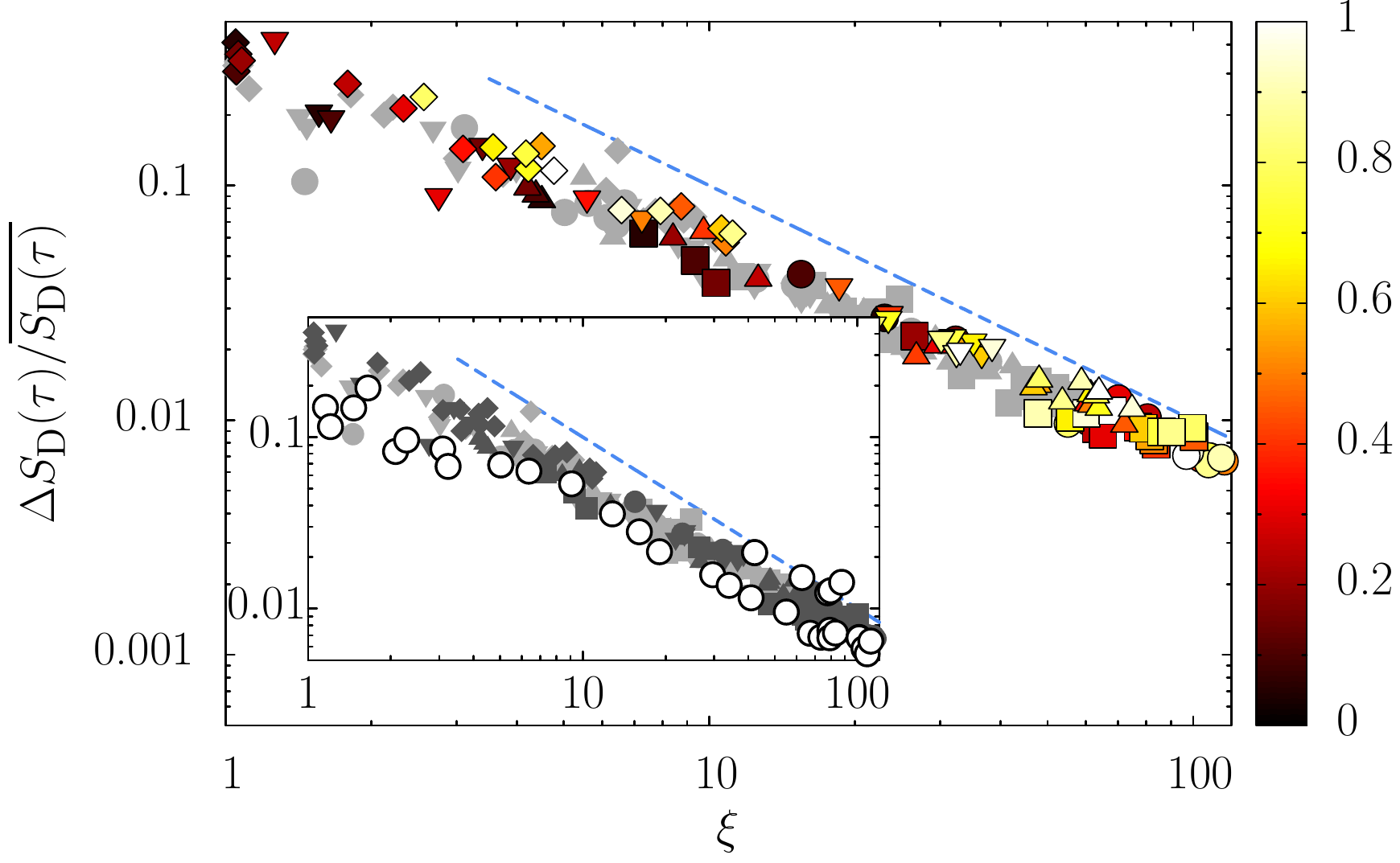} %{fluctuaciones2} 
\caption{(Color online) Main: Fluctuations $\Delta \SD/\langle \overline{\SD(\tau)}\rangle$ as a function of $\NPC$ for the 
DM with  same parameter values as Fig.~\ref{fig:SdeIPR}. The color-box indicates the values of $\lambda_0$. 
The dashed blue line is $1/\NPC$. Inset: fluctuations for the spin model ({\Large$\circ$}); the gray symbols 
correspond to the DM results (light gray $j=10$ ; dark gray $j=20$ ).
\label{fig:fluct}  
}
\end{figure}
%%%%%%%%%%%%%%%%%%%%%%%%%%%%%
In the main panel of Fig.~\ref{fig:SdeIPR} we show $\Sdec-\overline{\SD(\tau)}$  as a function of $\NPC$ for 
the DM with $\delta \lambda=0.1$ (the time window is $[\tau_0,\tau_0+250]$ with $\tau_0=10^7$). 
It is remarkable that all the data from Fig.~\ref{fig:lambda} collapse into a single curve for all energies and different values of $\lambda_0$, represented by different symbols colors.  It 
also tell us 
that when $\overline{\SD}$ attains the bound, 
the fluctuations become very small implying equilibration of the d-entropy.
It is clear that, as shown in Fig.~\ref{fig:lambda}, for large enough energies equilibration takes place below the transition 
value ($\lambdac$ or $\lambdac-\delta\lambda=0.4$).  A heuristic deduction of the universal curve can be obtained by looking at 
the extreme values. Large $\NPC$ values imply complete delocalization and therefore equilibration. 
In that case we see that the quantity $\Sdec-\overline{\SD(\tau)}$ tends to the universal value $(1-\gamma)$. 
In the opposite limit, 
when the initial state is more localized,
then $\Sdec\sim \overline{\SD(\tau)}$ with large fluctuations. 
The function that has both these limits and approximates well the numerical data is $(1-\gamma)(\NPC-1)/(\NPC+1)$.
In Fig.~\ref{fig:fluct}, we show that the fluctuations $\Delta \SD/\overline{\SD}$ behave as expected. They tend to 
decrease as the system becomes more delocalized. 
The deep relation between fluctuations -- albeit of a different quantity -- and the IPR was already 
studied and established in \cite{Marquardt2012}.

In order to test the generality of our results we consider another model.
A one-dimensional system of spin $1/2$ particles that interact through nearest-neighbor (NN) couplings. 
The quench is implemented by introducing next-nearest-neighbor (NNN) couplings   \cite{Santos2012}. 
The Hamiltonian of this spin model (SM) is given by 
\begin{eqnarray}
&& H(\lambda)   = H_0+\lambda V\\
&& H_0 = \sum_{i=1}^{L-1} J(S_i^x S_{i+1}^x +S_i^y S_{i+1}^y +\mu S_i^z S_{i+1}^z)\\
&& V	= \sum_{i=0}^{L-2} J(S_i^x S_{i+2}^x +S_i^y S_{i+2}^y +\mu S_i^z S_{i+2}^z).
\end{eqnarray}
where  $L$ is the number of particles and $S_i^{x,y,z}=\sigma_i^{x,y,z}/2$ are the spin 
operators, with $\sigma_i^{x,y,z}$ the 
corresponding Pauli matrices. The main difference with the DM, is that this is a 
quantum many-body system that lacks a semiclassical equivalent. 
The parameter $\lambda$ gives the NNN exchange with respect to the NN 
interaction in $H_0$.  We have considered a subspace with $L/3$ up spins and we fix $
\mu=0.5$ and $L=15$. 
Taking into account even parity the effective dimension is 
$N\sim (1/2)L! /[(L/3)!(L-L/3)!]$.
A remarkable feature of this system is that for $\lambda > 0.5$, like the DM, the 
system starts showing 
signatures of quantum chaos (recognizable in the change of level spacing statistics) 
\cite{Santos2012}.

We have checked (data not shown) that the behavior of  $\SD(\tau)$, $\Sdec-
\overline{\SD(\tau)}$ and the 
fluctuations as a function of $\lambda_0$ are equivalent to those shown in Fig.~
\ref{fig:lambda} for the DM.
Remarkably, in Fig.~\ref{fig:SdeIPR} one can observe that
the data obtained for the for the SM superimposes almost perfectly with the 
results obtained with the DM (and the conjectured curve).  Additionally,
the fluctuations behave like $1/\NPC$ for both the DM and the SM as it can be observed in  
Fig.~\ref{fig:fluct}.
The agreement is remarkable taking into account that both models differ significantly. 
We surmise that the IPR is the relevant figure of merit independently of the model and that 
the curve obtained can be conjectured to be universal.

%%%%%%%%&&&&&&&&&&&&&&&&&&&&&&&&&&&&&&&&&&&&&&&
\begin{figure} %[htb]
\includegraphics[width=1.0\linewidth]{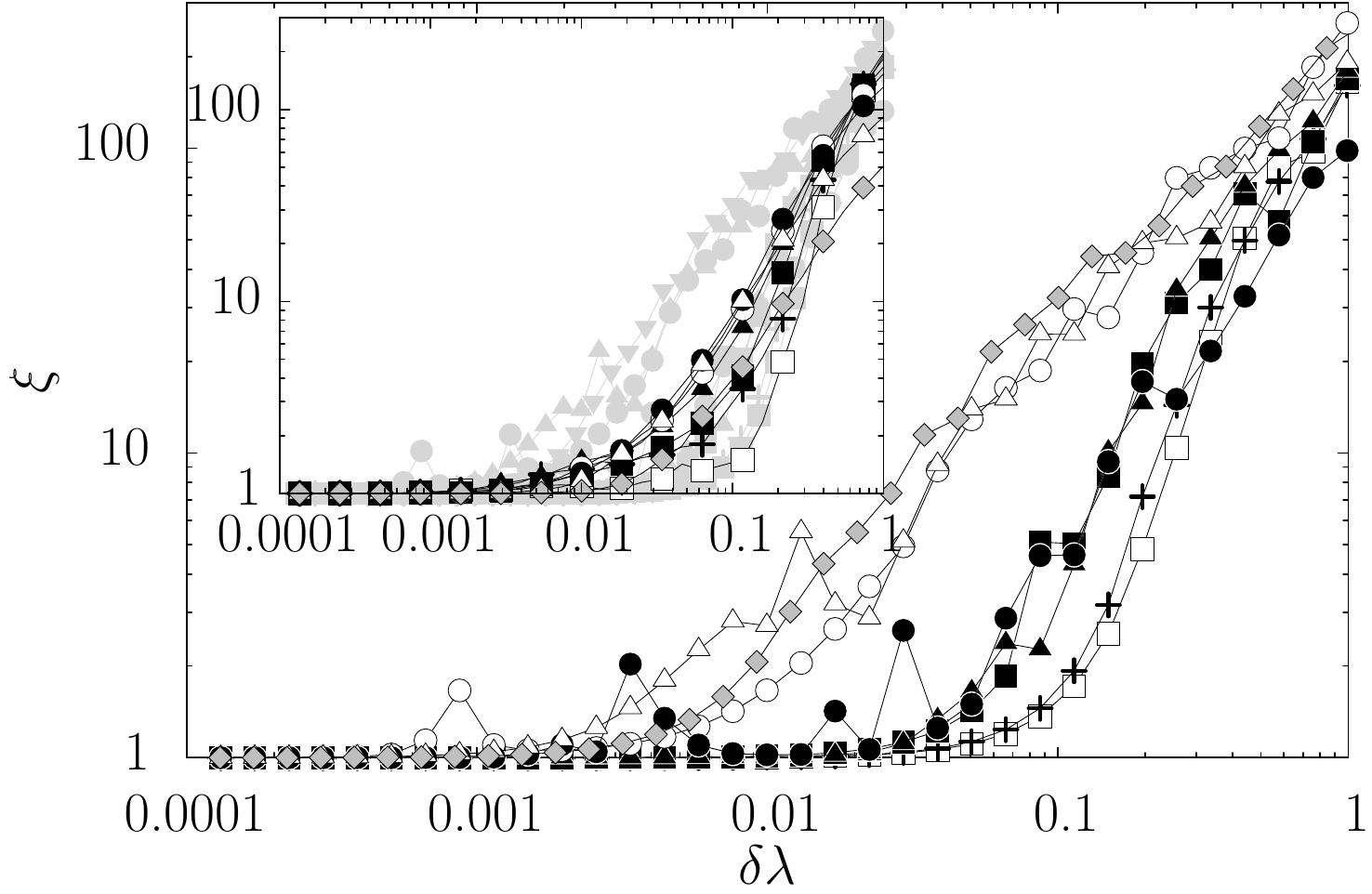} %%{IPR_DeltaLam_entorno100} 
\caption{\label{fig:IPRdelam} $\NPC$ (IPR)  as a function of $\delta\lambda$
for the DM with $j=20$ ($\nmax=200$), initial state $\ket{100}$, and different values of $
\lambda_0$. 
In the inset we show the same for the spin model with $\mu=0.5$. 
In both cases the values of $\lambda_0$ are: ($\square$) 0.01,
 ($\pmb{+}$) 0.1, ($\blacksquare$) 0.2, ($\blacktriangle$) 0.3, ({\Large $\circ$}) 0.4, ({\Large $\bullet$}) 0.5, ($\triangle$) 0.7, ($\diamond$) 1.0. The light-gray symbols in the inset correspond to the ones in the main panel (``eye guide'').
 } 
\end{figure}
%%%%%%%%%%%%%%%%

We have shown -- through the IPR -- that the equilibration process depends on the spreading of 
the initial state over the perturbed basis rather than on the quantum chaos parameter.
A related question that arises is, 
what is the relation, if any, between the IPR and the transition form integrability to quantum chaos 
\cite{Georgeot1997,Marquardt2012}. For this reason, in Fig.~\ref{fig:IPRdelam} we show $\NPC$ 
as a function of the quench $\delta \lambda$ for the DM (and the SM in the inset), and 
different values of the coupling 
parameter $\lambda_0$. 
Three different regimes can be distinguished \footnote{The IPR is directly related to the local density of states (LDOS). We identify the three regimes with those described in \cite{Wisniacki2013_2} for the LDOS. For very small $\delta \lambda$ perturbation theory applies, and the LDOS is approximately a delta function. For intermediate values, the LDOS is a Lorentzian. We consider large $\delta \lambda$ the regime where the LDOS takes a much more complicated shape, in our case for $\delta\lambda\approx 0.1$}.
For very small $\delta \lambda$, $\NPC$ remains very close to its initial value (unity). Actually looking at $(\NPC-1)$ 
a quadratic regime is observed (not shown), which is expected from perturbation theory.
On the other hand, for large values of $\delta\lambda$, the $\NPC$ saturates, and the curves for all coupling values collapse.
The subtle behavior occurs for the small-to-intermediate values of $\delta\lambda$.
We see in the chaotic regime, $\NPC$ starts  to grow for much smaller values. However, the slope in the power law is smaller than for the 
integrable regime. On the other hand, the integrable case, needs a much stronger quench to trigger the power law behavior. 
This behavior is not as clearly observed in the spin chain. This is probably related to the absence of a semiclassical limit 
for this system. However, the definitive connection between the classical dynamics and the IPR remains to be understood 
and we leave the discussion for a future work \cite{Unpub2015}.
 
To conclude, using the fact that the relaxation to equilibrium is reflected 
in the saturation of the 
sub-extensive correction to the mean d-entropy which is correlated with decreasing small 
fluctuations, we have shown that equilibration  -- in a cyclic process consisting of two 
instantaneous quenches -- depends on the energy localization properties of the initial state, 
and not necessarily in the degree of chaoticity of the evolution. 
Our numerics were done using two systems 
which undergo a transition from integrable to chaotic, but they are fundamentally different  since
one of them does not have a semiclassical counterpart. We have found that for a wide range of parameters, 
there is a clear functional relation between equilibration (saturation of d-entropy) and localization (via the IPR).
This, and the fact that the systems are fundamentally different, provides strong evidence that there is a 
universal relation between localization and relaxation, besides integrability. 

%\noindent \textit{Acknowledgments.}
I.G.M. and D.A.W.
received support from ANPCyT (Grant No.PICT 2010-1556), UBACyT,
and CONICET (Grants No. PIP 114-20110100048 and No. PIP 11220080100728).
 A.J.R. acknowledges support from ANPCyT (Grants No. PICT-2010- 02483 and No. PICT-2013-0621), CONICET, and UBACyT.
 %%%%%%%%%%%%%%%%%%%%%%%%%%%%
%%%%% commentar las tres lineas siguientes para incluir la bibliografia que esta debajo
%\bibliographystyle{apsrev4-1}
%\bibliography{refs}
%\end{document}
\end{document}